\documentclass[twocolumn,english,aps,prb,showpacs,byrevtex,amsmath,amssymb,superscriptaddress]{revtex4-1}
\usepackage[T1]{fontenc}
\usepackage[latin9]{inputenc}
\usepackage{textcomp}
\usepackage{amsmath}
\usepackage{graphicx}
\usepackage{multirow}

\makeatletter
\usepackage{lipsum}
\usepackage{epsfig}
\usepackage{subfigure}
\usepackage{color}
\usepackage{physics}

\definecolor{Red}{rgb}{1,0,0}
\definecolor{Blu}{rgb}{0,0,1}
\definecolor{Green}{rgb}{0,1,0}

\makeatother

\usepackage{babel}
\begin{document}

\title{Topological transition in Pb$_{1-x}$Sn$_{x}$Se using Meta-GGA}

\author{Rajibul Islam}
\email{Rajibul.Islam@magtop.ifpan.edu.pl.}
\affiliation{International Research Centre Magtop, Institute of Physics, Polish Academy of Sciences,
Aleja Lotnik\'ow 32/46, PL-02668 Warsaw, Poland}

\author{Giuseppe Cuono}

\affiliation{Dipartimento di Fisica "E.R. Caianiello", Universit\`a degli Studi di Salerno, I-84084 Fisciano
(SA), Italy}

\affiliation{International Research Centre Magtop, Institute of Physics, Polish Academy of Sciences,
Aleja Lotnik\'ow 32/46, PL-02668 Warsaw, Poland}

\author{Nguyen Minh Nguyen}

\affiliation{International Research Centre Magtop, Institute of Physics, Polish Academy of Sciences,
Aleja Lotnik\'ow 32/46, PL-02668 Warsaw, Poland}

\author{Canio Noce}
\affiliation{Dipartimento di Fisica "E.R. Caianiello", Universit\`a degli Studi di Salerno, I-84084 Fisciano
(SA), Italy}
\affiliation{Consiglio Nazionale delle Ricerche CNR-SPIN, UOS Salerno, I-84084 Fisciano (Salerno),
Italy}

\author{Carmine Autieri}
\affiliation{International Research Centre Magtop, Institute of Physics, Polish Academy of Sciences,
Aleja Lotnik\'ow 32/46, PL-02668 Warsaw, Poland}
\affiliation{Consiglio Nazionale delle Ricerche CNR-SPIN, UOS Salerno, I-84084 Fisciano (Salerno),
Italy}

\date{\today}
\begin{abstract}
		
We calculate the  mirror Chern number (MCN) and the band gap for the alloy Pb$_{1-x}$Sn$_{x}$Se as a function of the concentration x by using virtual crystalline approximation.
We use the electronic structure from the relativistic density functional theory calculations in the  Generalized-Gradient-Approximation (GGA) and meta-GGA approximation. Using the modified Becke-Johnson meta-GGA functional, our results are comparable with the available experimental data for the MCN as well as for the band gap.
We advise to use modified Becke-Johnson approximation with the parameter $c$=1.10 to describe the transition from trivial to topological phase for this class of compounds.

\end{abstract}

\maketitle

\section{Introduction}

Rock-salt chalcogenides is a large class of functional materials, which exhibits many novel properties such as superconductivity\cite{matsushita2006type}, thermoelectricity\cite{wood1988materials,wang2011heavily}, and ferroelectricity\cite{lebedev1994ferroelectric}, with interesting applications in the optoelectronics\cite{akimov1993carrier,liu2010dependence} and spintronics\cite{jin2009large,grabecki2007quantum}.
These materials attracted considerable attention also because of the prediction\cite{fu2011topological} and realization\cite{hsieh2012topological} of the topological crystalline insulating phase in SnTe. The surface bands of the topological crystalline insulators (TCI) are low-energy states occurring along certain high-symmetry directions.\cite{Barone2013,Silvia2013,Volobuev17}  The nature of these low-energy surface states is sensitive to the surface orientation, and unlike the surfaces states in the conventional Z$_2$ topological insulators \cite{fu2007topological}, the surface states of TCI are protected by crystalline symmetry, instead of time reversal symmetry. 
We point out that also the topological crystalline metallic phases have been proposed to be present in orthorhombic structures\cite{Chen2015NatComm,Daido2019,Cuono2019,AutieriNoce2017PhilMag,Autieri2017JPCM}.

Moreover, TCIs are not yet detected in compounds of the group IV-VI, semiconductors with a crystal structure similar to rock-salt chalcogenides\cite{Nimtz1983,khokhlov2002lead}, like PbTe and PbSe, so that
their substitutional alloys, such as Pb$_{1-x}$Sn$_x$Te and Pb$_{1-x}$Sn$_x$Se, may represent an interesting  possibility to investigate the transition between topologically trivial and topologically nontrivial phase\cite{dziawa2012topological,tanaka2012experimental,xu2012observation,wojek2014band}.
Indeed, interfacing or alloying  different materials can produce exotic phases or new properties that are not present in the bulk\cite{Paul2014APL,Roy2016SciRep,Autieri2012PRB}. 
We also note that in the case of the rock-salt chalcogenides alloys, it was proposed the presence of a Weyl phase\cite{Lusakowski2018PRB,Wang2019} in the intermediate region between the trivial and the TCI phase.

In this paper, we use the Virtual Crystal Approximation (VCA) to estimate the transition between topologically trivial and non-trivial phases, increasing the content of Sn in the Pb$_{1-x}$Sn$_x$Se alloy. Our goal is to look into the concentration of Sn which leads to the transition from a topologically trivial phase to a non-trivial phase in Pb$_{1-x}$Sn$_x$Se alloy, applying the Density Functional Theory (DFT). We recall that the experimental transition from direct to the inverted band structure is observed for $x>0.16$\cite{Nimtz1983}. We look also to the role played by the spin-orbit coupling (SOC) in driving this transition. We stress that SOC is a relativistic effect that may give rise to exotic effects like complex spin texture\cite{Autieri2019}, spin Hall effect\cite{Slawinska20192D}, topological phases\cite{Groenendijk2019}, large magnetocrystalline anisotropy\cite{Ming2017PRB,Autieri2014NJP} and non collinear phases \cite{Ivanov2016InorgChem}.

The paper is organized in the following way: in section II we describe the computational techniques adopted; Section III contains the results concerning the bulk band structures of PbTe, PbSe and SnTe as well as the mirror Chern number calculation (MCN) of Pb$_{1-x}$Sn$_x$Se alloy, while in Section IV we summarize the main results of the paper.

\begin{figure}[t!]
\centering
\includegraphics[width=6.3cm,angle=270]{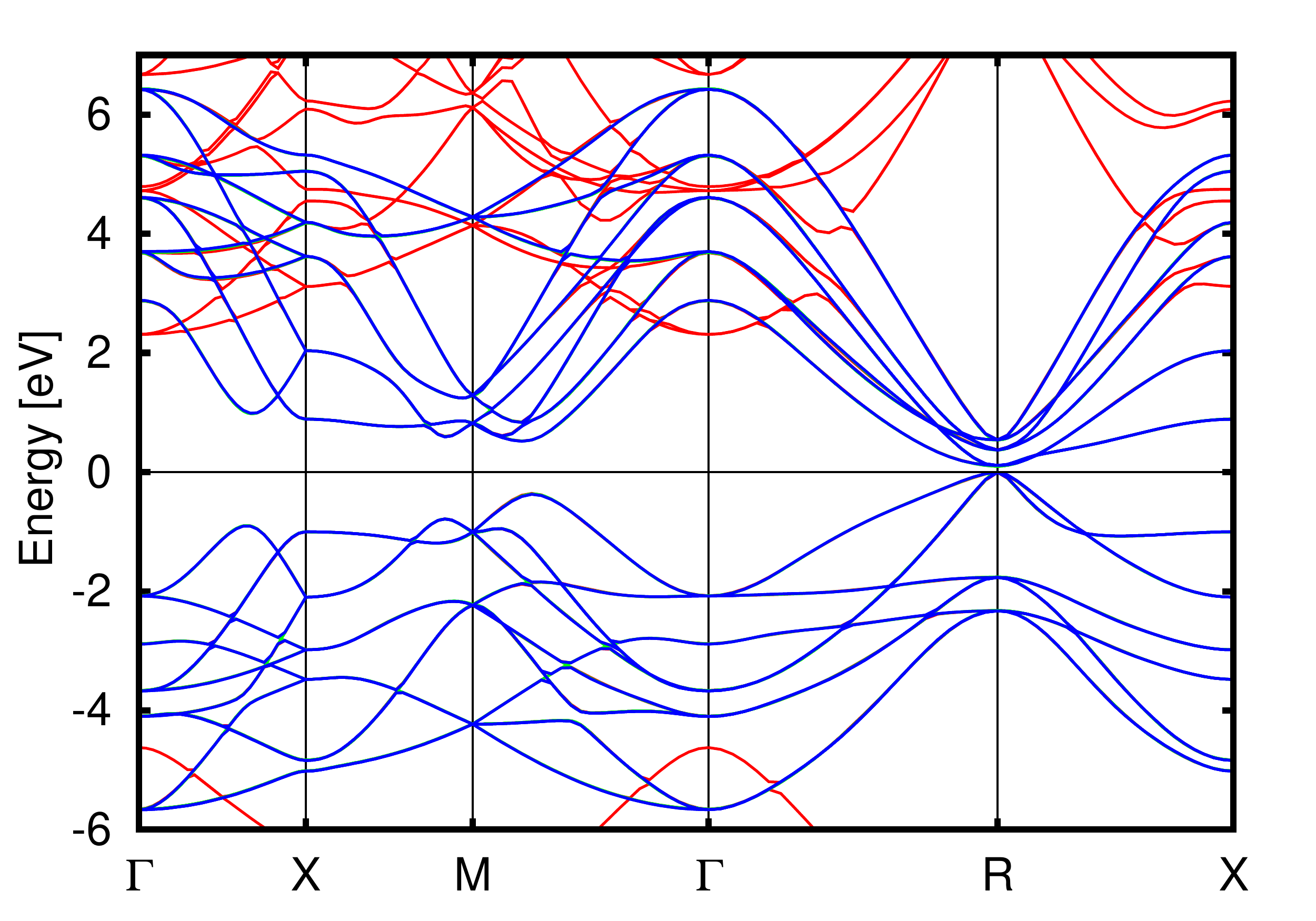}
\caption{Band structure of the SnTe along high symmetry paths of the cubic Brillouin zone.
MBJLDA band structure (red) and  band structure obtained considering the p-like Wannier functions (blue).
The Fermi level is set at the zero energy.
}
\label{SnTe}
\end{figure}

\section{Computational details}

We have performed relativistic first-principles DFT calculations by using the VASP package\cite{VASP_package} based on plane wave basis set and projector augmented wave method
\cite{pseudo}. A plane-wave energy cut-off of 250~eV has been used. We have performed the calculations using 6$\times$6$\times$6 k-points centred in $\Gamma$ with 216 k-points in the independent Brillouin zone. 
The band gap is underestimated in DFT, so that several methods have been applied with the aim to increase the band gap in DFT calculations \cite{Keshavarz2017PRB,Autieri2016JPCM}. Also in this class of compounds the GGA approximation underestimates the band gap\cite{Goyal17}. If we use the GGA approximation we do not obtain the topological transition since we get topological compounds for every concentration at the volume considered. To avoid this problem, recently it was used the GGA+U with the Coulomb repulsion on the 6s of the Pb.\cite{Wang2019}
Differently from the literature, the modified Becke-Johnson exchange potential together with local density approximation for the correlation potential scheme MBJLDA\cite{MBJLDA1,MBJLDA2} has been considered in this work.
We point out that MBJLDA improves the description of the band gap\cite{Camargo12} that is crucial for the topological properties.

Moreover, after obtaining the Bloch wave functions\cite{Marzari97,Souza01} $\psi_{n,\textbf{k}}$, the p-like anion and cation Wannier functions are build-up using the WANNIER90 code\cite{Mostofi08}. To determine the real space Hamiltonian in the Wannier function basis, we have used the Slater-Koster interpolation scheme, and  we have constructed the symmetrized relativistic Wannier tight-binding model using an independent python package wannhrsymm\cite{WANNSYM}. 

Since the DFT is a zero temperature theory, in the calculations we have considered the room temperature lattice constants\cite{McCann87} reduced by 0.5\% to take in account the temperature effect. The values used are: $a_{SnTe}$=6.2964~\AA, $a_{PbTe}$=6.4277~\AA ~  and $a_{PbSe}$=6.0954~\AA.

\section{Results and discussion}

\begin{figure}[t!]
\centering
\includegraphics[width=6.3cm,angle=270]{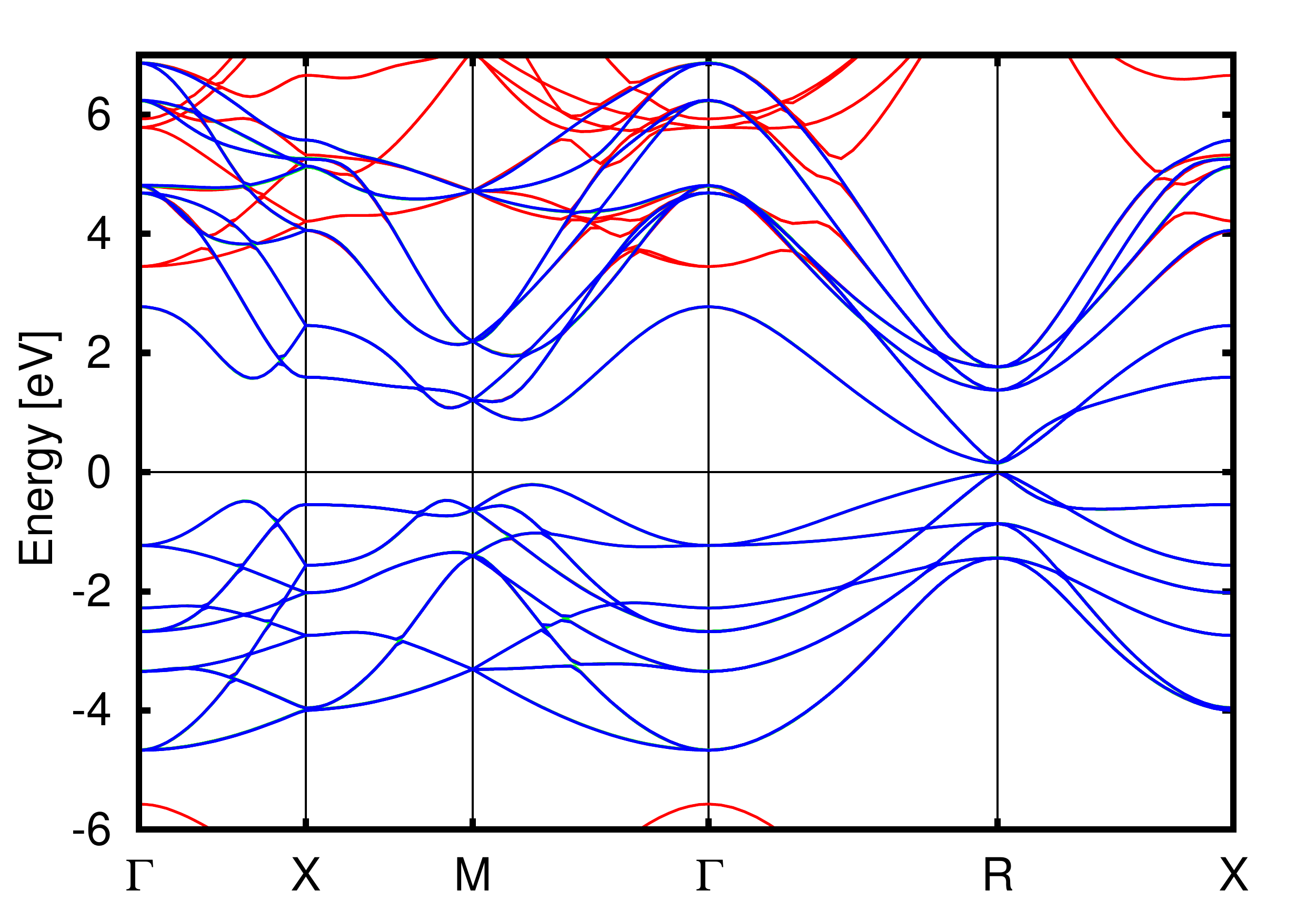}
\caption{Band structure of the PbTe along high symmetry paths of the cubic Brillouin zone.
MBJLDA band structure (red) and  band structure obtained considering the p-like Wannier functions (blue).
The Fermi level is set at the zero energy.
}
\label{PbTe}
\end{figure}

\begin{figure}[t!]
\centering
\includegraphics[width=6.3cm,angle=270]{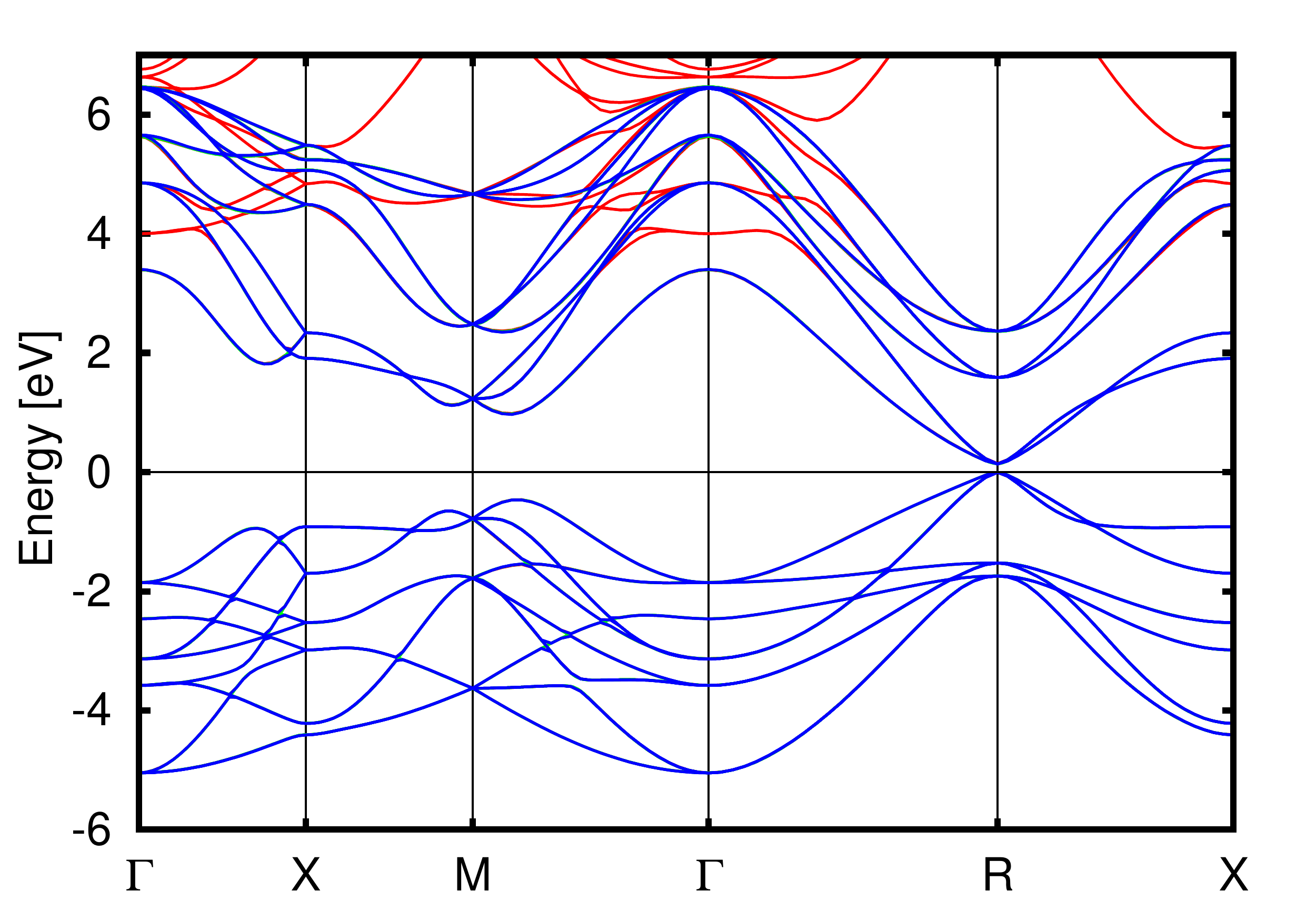}
\caption{Band structure of the PbSe along the high symmetry paths of the cubic Brillouin zone.
MBJLDA band structure (red) and  band spectrum obtained considering the p-like Wannier functions (blue).
The Fermi level is set at the zero energy.
}
\label{PbSe}
\end{figure}

\noindent To calculate the MCN, we choose a unit cell with rock-salt structure containing $2\times2\times2$ atoms in the three directions. The low-energy physics is dominated by the $p$-anion and $p$-cation orbitals so that we have 48 bands, 24 of which are occupied and 24 unoccupied.
In Figs.~\ref{SnTe}, ~\ref{PbTe} and ~\ref{PbSe} we report the relativistic band structures of SnTe, PbTe and PbSe, respectively. The bandwidth is about 12 eV and the fundamental gap of the bands is located at the R point ($\pi,\pi,\pi$) of the Brillouin zone.

Table \ref{tab1} shows the on-site energies  $\epsilon$, the first, the second and the third neighbour hopping parameters and the spin-orbit coupling $\lambda$ constants for SnTe, PbTe and PbSe.
We denote by $t^{lmn}_{\alpha,\beta}$ the hopping amplitudes along the connecting direction l$\mathbf{x}$ + m$\mathbf{y}$ + n$\mathbf{z}$ 
between the orbitals $\alpha$ and $\beta$.
Where it is possible, we indicate also the $\sigma$ and $\pi$ chemical bonds.
The first neighbour hopping is the cation-anion hopping; the second and third ones are anion-anion and cation-cation hopping terms. 
From the data here reported, we infer that for SnTe the on-site energies of the cation and the anion are almost symmetrical with respect to the Fermi level, differently from the PbTe and PbSe.
In the case of SnTe, the hoppings reported in the Table are sufficient to reproduce the topological behaviour.
If we assume $t^{200}$=0, the system becomes trivial. 
Considering the SnTe compound, the Te-$s$ and the Sn-$d$ hybridizations produce a variation of the sign of the Sn-Sn second neighbour hopping $t^{110}_{\sigma}$ with respect to the Te-Te second neighbour hopping $t^{110}_{\sigma}$.  

\begin{table} [!]
\begin{center}
\begin{tabular}{ |c||c||c|c|c|  }
 \hline
& & SnTe & PbTe  & PbSe \\
 \hline
\multirow{2}{4em}{On-site} & $\epsilon_{c}$  &  1094.6     & 1855.4  &  2250.4  \\
& $\epsilon_{a}$  & -1205.4  & -646.0    & -1206.1   \\
\hline
 \multirow{2}{4em}{SOC} & $\lambda_{c}$ & 334.5 & 1068.2  & 1132.5  \\
&  $\lambda_{a}$    & 524.4 & 553.8 &  250.3  \\
 \hline
 \multirow{2}{4em}{cation-anion} &   $t^{100}_{px,px}$=$t^{100}_{\sigma}$   & 1916.1 & 1807.4 & 1788.9\\
 & $t^{100}_{py,py}$=$t^{100}_{\pi}$  & -434.2   & -402.6 & -353.6\\
  \hline
 \multirow{4}{4em}{cation-cation} & $t^{200}_{px,px}$=$t^{200}_{\sigma}$  &     281.7 & 268.4   & 311.8  \\
 & $t^{200}_{py,py}$=$t^{200}_{\pi}$  &     82.9 & 81.6   & 96.1  \\
 & $t^{110}_{px,px}$  &     -12.1 & 2.9   & 33.0 \\
 & $t^{110}_{pz,pz}$  &     9.4 & 5.3   & -1.9 \\
 & $t^{110}_{px,py}$  &     13.2 & 26.7   & 11.9 \\
 \hline
 \multirow{4}{4em}{anion-anion} & $t^{200}_{px,px}$=$t^{200}_{\sigma}$  &     -21.8 &  12.9  & -87.9 \\
 & $t^{200}_{py,py}$=$t^{200}_{\pi}$  &     34.1 & 23.5  & 14.2 \\
 & $t^{110}_{px,px}$ &     23.9 & 20.3  & 47.9 \\
 & $t^{110}_{pz,pz}$  &     -10.9 &  -12.9 & -24.3 \\
  & $t^{110}_{px,py}$  &     -224.7 & -111.2   & -156.1 \\
 \hline
\end{tabular}
\end{center}
\caption{Values of the electronic parameters of our tight-binding model for SnTe, PbTe and PbSe. The unit is meV.}
 \label{tab1}
\end{table} 

\noindent Then, we analyse the trend of the MCN and the band gap for the ternary compound Pb$_{1-x}$Sn$_{x}$Se as a function of the concentration $x$, and we study the transition from the trivial to the topological phase as a function of $x$.

The VCA Hamiltonian that describes the alloy is the following:
\begin{equation}
\label{eqn:alloy}
H(x)=H_{PbSe} + x(H_{SnTe}-H_{PbTe})\, ,
\end{equation}
where $H_{PbTe}$, $H_{PbSe}$ and $H_{SnTe}$ are the symmetrized relativistic Wannier tight-binding Hamiltonians for the PbTe, PbSe and SnTe compounds, respectively.
In this model, the hopping parameters are the outcome of the DFT calculations.
In this analysis, we exclude the SnSe compound for two reasons: it is far from the transition and has a rhombohedral structure instead of a cubic one, that is necessary to fulfil the symmetries for the MCN. 

We note that the Hamiltonian exhibits the mirror and the time-reversal symmetries.
The mirror operator respect to the ($\overline{1}$10) plane is the following:
\begin{equation}
\label{eqn:mirror}
M_{xy}=\frac{1}{\sqrt{2}}(\sigma_{x}-\sigma_{y})\otimes(\mathbb{1}_{3}-L_{z}^{2}-\comm{L_{x}}{L_{y}})\otimes P_{xy}\, .
\end{equation}
It is a Kronecker product of three parts, the first is related to the spin, the second to the orbital and the third to the  atomic degrees of freedom.
$\sigma_{x}$ and $\sigma_{y}$ are the Pauli matrices, $\mathbb{1}_{n}$ is the n$\times$n identity matrix, $L_{i}$ with $i=x,y,z$ are the orbital angular momentum operators for $l=1$, and $P_{xy}$ is the matrix that exchanges the atomic positions respect to the mirror plane, whose form depends on our choice of the vector basis.
The time-reversal operator is $T=\sigma_{y}\otimes \mathbb{1}_{3}\otimes \mathbb{1}_{8},$ where also in this case the first part is related to the spin, the second to the orbital and the third to the atoms.
These operators verify the following relations:
\begin{equation}
\label{eqn:actionmirror}
M_{xy}H(k_{x},k_{y},k_{z})M^{\dagger}_{xy}=H(k_{y},k_{x},k_{z})\, ,
\end{equation}
\begin{equation}
\label{eqn:actiontime}
TH(k_{x},k_{y},k_{z}){T}^{\dagger}=H(-k_{x},-k_{y},-k_{z})\, ,
\end{equation}
\noindent and the mirror operator anticommutes with the time-reversal operator. In the eigenbasis of the mirror operator, the time-reversal takes off-diagonal block and the Hamiltonian commutes with the mirror operator. Therefore, it has a block-diagonal form. This leads the MCN vanishing on the high symmetry planes, where each block has an opposite MCN.

\begin{figure}[t!]
\centering
\includegraphics[width=8.5cm, angle=0]{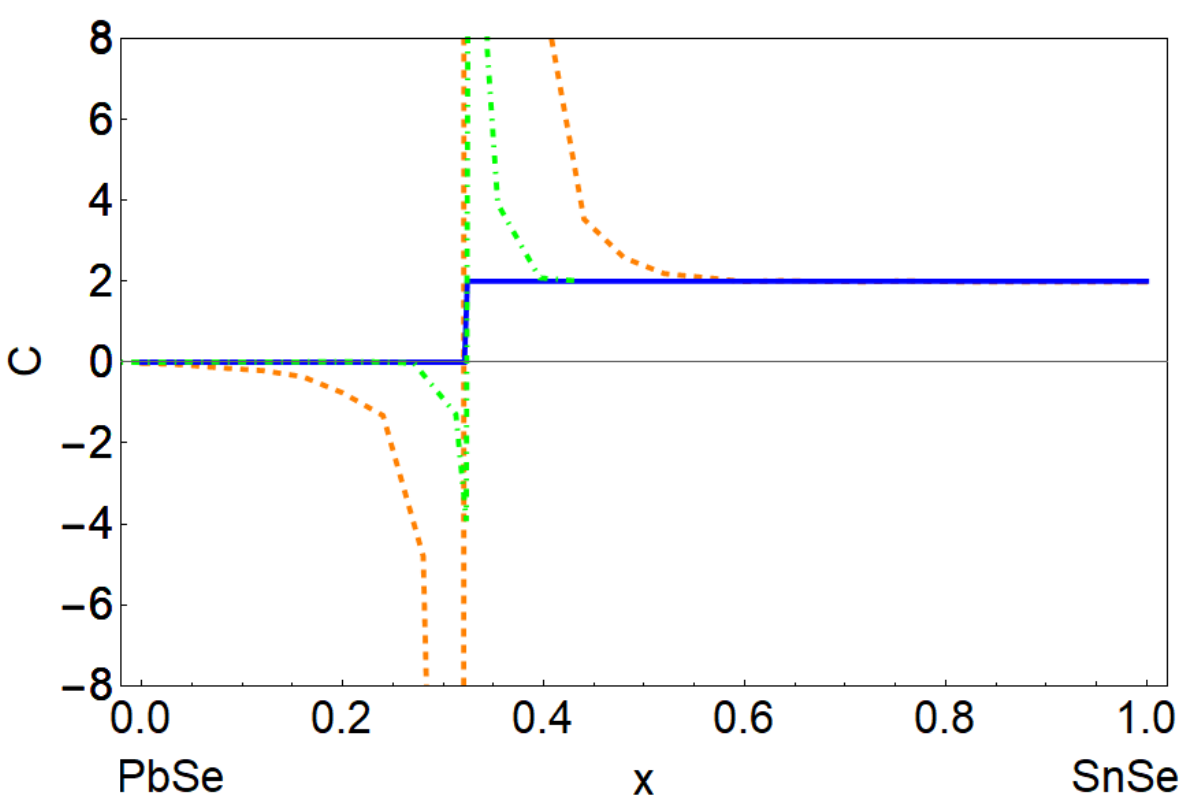}
\caption{MCN as a function of the $x$ composition for the Pb$_{1-x}$Sn$_{x}$Se alloy obtained with different grids of integration points. The orange dashed line is the result with 40 integration points, the green dot-dashed line with 200 integration points and the blue solid line is the ideal result. 
}
\label{Chernkk}
\end{figure}

To calculate the MCN we use the Kubo formula:
\begin{equation}
\label{chernnumber}
\footnotesize C=\frac{1}{\pi}\int_{0}^{2\pi}\int_{0}^{2\pi}\sum_{n\leq n_{F}, n'>n_{F} } Im\left[\frac{\small \langle n \small \lvert \partial_{k_{x}}H \rvert n' \small  \rangle \small \langle n' \small \lvert \partial_{k_{z}}H  \rvert  n \small \rangle}{(\epsilon_{n}-\epsilon_{n'})^2}\right]dk_{x}dk_{z}\, ,
\end{equation}
where $ \rvert n \small \rangle $ and $\epsilon_{n}$ are the eigenstates and the eigenvalues of the Hamiltonian in the projected subspace and $n_{F}$ is the filling. For our unit cell, we have  $n$=24 and $n_{F}$=12 for all compounds. 
When the MCN is equal to 0 we have a trivial insulator, while when is equal to 2 we have a TCI.
We have tested the numerical convergence of the integral reported in Eq. (\ref{chernnumber}) as a function of the integration points in the k$_x$k$_z$-space.
In Fig. \ref{Chernkk} we plot the MCN as a function of the doping $x$ obtained with a number of integration points equal to 40 and 200, and the ideal result, when this number goes to infinity.
When the number of integration points is too low and we are close to the topological transition, the integration fails since it gives large Chern numbers, while, by increasing the number of integration points, the integral slowly converges. 

\begin{figure}[t!]
\centering
\includegraphics[width=8.5cm, angle=0]{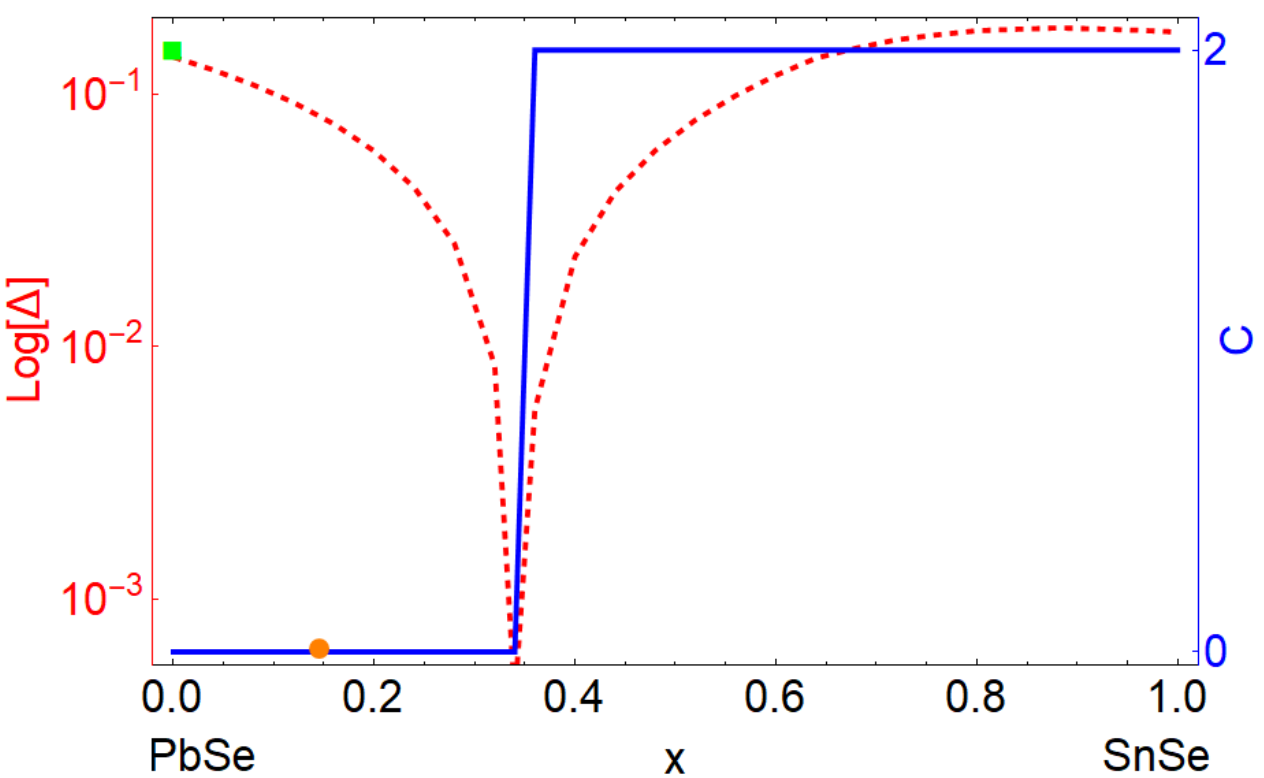}
\caption{MCN (solid blue line)and band gap (dashed red line) as a function of the concentration $x$ for the Pb$_{1-x}$Sn$_{x}$Se alloy. The band gap is plotted in logarithmic scale. The green square indicates the experimental band gap for PbSe while the orange circle represents the experimental point of the transition from the trivial to the topological phase. 
}
\label{Chernmap}
\end{figure}

In Fig. \ref{Chernmap} we plot the MCN and the band gap as a function of the concentration $x$ and we compare our results with the available experimental data. We find that the PbSe is a trivial insulator while the SnSe is a topological crystalline insulator. Furthermore, the critical point where the transition from the trivial to the topological phase is located is at $x$=0.33, even though the experimental critical concentration\cite{Nimtz1983} is found at $x$=0.16. At this point, we also find that band gap goes to zero. The experimental band gap\cite{Nimtz1983} for PbSe is 145 meV , while in our case we get 139 meV, whereas for the hypothetical rock-salt SnSe structure is 176 meV.

For these calculations we used the MBJLDA with the parameter $c$=1.10 that gives us a reasonable results for both the band gap and the topological transition.
A larger value of $c$ would improve the agreement with the experimental gap even though the topological transition would not be very well described; on the other hand, a smaller value of $c$ would produce an opposite trend.
Indeed, increasing $c$ the critical concentration increases too shrinking the topological region;  while reduced values for $c$ lead to an reduction of the critical concentration and consequently an enhancement of the topological region.

\section{Conclusions}

We have calculated the MCN and the band gap of the ternary compound Pb$_{1-x}$Sn$_{x}$Se as a function of the concentration $x$ in VCA approximation by using the symmetrized relativistic electronic structure.
We infer that the MBJLDA approximation gives a description of the experimental results better than GGA approach.
In particular, we underline that we need to use a low value of the $c$ parameter to obtain a good agreement with the available experimental data for both MCN and band gap. 
Finally, we would like to note that more accurate techniques like SQS\cite{Lusakowski2018PRB,Wang2019} are required to investigated the Weyl phase in DFT and more sophisticated methods are necessary for the MCN calculation in alloy\cite{Varjas2019}.

%\section*{Acknowledgments}
\vspace{5mm}

We would like to thank W. Brzezizki, T. Dietl, T. Hyart, J. S{\l}awi{\'{n}}ska and V. V. Volobuev for useful discussions. The work is supported by the Foundation for Polish Science through the IRA Programme co-financed by EU within SG OP. This research was carried out with the support of the Interdisciplinary Centre for Mathematical and Computational Modelling (ICM) University of Warsaw under Grant No. G73-23 and G75-10.

\bibliography{SnTe}

\end{document}